\documentclass[twocolumn,showpacs,preprintnumbers,amsmath,amssymb]{revtex4}
\usepackage{graphicx}

\begin{document}
\title{Vortex lattices: from the hydrodynamic to the quantum hall regime}
\author{L. O. Baksmaty, S.J. Woo, S. Choi, N. P. Bigelow}
\affiliation{Dept. of Physics and Astronomy, and Laboratory for Laser 
Energetics, University of Rochester, Rochester, NY 14620}
\date{\today}
\begin{abstract}
We describe the excitations of a vortex lattice in rapidly rotating trapped Bose-Einstein Condensates, focusing on their evolution from the hydrodynamic to the quantum Hall regimes. We observe the dramatic changes and show that single particle states are an increasingly dominant signature of the spectrum.  We interpret the behavior by analogy with the integral quantum Hall effect. Furthermore, we compare our results to recent experiments at JILA and find excellent agreement.

\end{abstract}
\pacs{03.75.Fi, 05.30.Jp, 42.50.Vk}\maketitle

\begin{figure}
\includegraphics*[width=6.5cm,height=2.0cm]{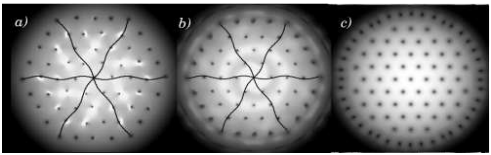}
\caption{ Azimuthally symmetric (3,0)~\cite{anglin} a) Tkachenko mode 
and b) inertial mode. c) Surface wave }
\label{display}
\end{figure}

The angular momentum of a rapidly rotating trapped Bose-Einstein condensate (BEC) is carried in 
singly quantized vortex lines. At equilibrium, the vortices typically form a triangular array which rotates 
with the condensate~\cite{sonin,anglin} -- the Abrikosov lattice.  
Recent experimental and theoretical efforts to understand the dynamics of this vortex lattice have focused on the calculation~\cite{anglin,baym,lobaks,nippon,cozzini,choi} and measurement~\cite{jilat,jilat2} 
of two particular branches of the spectrum commonly described in terms
of the associated Tkachenko and inertial modes.
The Tkachenko waves which are governed by the small but finite shear modulus ($C_{2}$), 
mainly involve fluctuations in the velocity field of the condensate and appear
to an observer as elliptically polarized transverse waves moving through the lattice
(Fig.~\ref{display}a). On the other hand inertial modes which mainly involve density fluctuations similar to 
hydrodynamic shape oscillations are characterized by the bulk modulus $(C_{1})$. 
The finite size of the vortices ensures that both modes are intimately coupled 
and they can be shown to occur as two spectral branches of the coupled 
fluid-lattice equations in a hydrodynamic 
treatment~\cite{baym}. 

In the BEC, the angular frequency of the trap rotation ($\Omega \hat{\bf z}$) plays the role analogous to a uniform magnetic field applied to a type-II superconductor or to a field effect transistor (FET) as in the integer quantum Hall effect (IQHE)~\cite{yoshi}. When investigating vortices in the trapped BEC, it is natural to ask whether the connection to the IQHE can provide insight and lead to new results. In this letter we concern ourselves with the transformation of the excitation spectrum as we 
transition from the stiff Thomas-Fermi (TF) regime to the mean field quantum Hall regime (QHR).  We highlight edge modes associated with the boundary of the condensate which appear as a ring of vortices precessing the apparently undisturbed ground state (Fig~\ref{display}c).
These modes are particle-like excitations that have an intimate connection to Landau levels in the IQHE and are shown to be especially important for the BEC as the QHR is approached. Our identification evokes an interpretation of the Tkachenko and inertial modes as analogues of extended and bound single particle states fundamental to the IQHE. Finally, we use this framework to discuss lattice meting and vortex nucleation.

One exciting feature of trapped BECs is the wide range of rotational speeds over which vortex lattices may be produced.  As $\Omega$ increases from 0 to the  radial trapping freqency $\omega_{\bot}$, centrifugal effects cause the particle density to change and
we encounter three physical regimes~\cite{baym} which may be identified by three 
length scales defining the lattice. These are the radius of the condensate in the $x-y$ plane ($R$), the 
healing length($\xi$) 
and the intervortex spacing ($\epsilon$) defined by $\xi=1/\sqrt{8\pi na}$ and 
$\epsilon^{2}=2\pi \hbar / (\sqrt{3} m\Omega)$. Here $a$ and $n$ represent the s-wave scattering length of the trapped species and density of particles at the trap center.
 $\epsilon$ and $\xi$ are the coherence lengths
of Tkachenko and inertial modes and togther with the wave vector $\vec{k}$ they determine 
whether an excitation has a collective or particle-like character. 
In the stiff $(\epsilon^2 \! \gg \! \xi \! R)$ TF regime $(\xi \! \ll \! \epsilon \! \ll \! R )$, which 
occurs at slow rotation, 
the time scale of density fluctuations is much shorter than that of rotation
and the system responds to perturbations as a rigid body. At faster speeds we encounter the 
soft ($\epsilon^{2} \ll \xi R $) TF regime  
indicating that while the mean field interaction continues
to dominate the properties of the lattice, the dynamics of the density had become more 
sluggish and thus important in the response of the lattice to perturbations. At ultra-fast
rotation we enter the QHR ($\epsilon \! < \! \xi \! < \! R $) , here the chemical potential 
$\mu$ is dominated by the cyclotron energy.
In the QHR, the lattice is described by the primary occupation of the almost degenerate 
lowest Landau level (LLL)~\cite{ho} and is expected to undergo quantum phase transitions into a myriad
of strongly correlated vortex liquid states~\cite{wilkin}. The change in character of the 
excitations as a function of the relative sizes of the length scales ($\xi, \epsilon, R$)
is an important theme in the transformation of the spectrum and the subject of this paper.
\begin{figure*}
\includegraphics*[width=16.5cm, height=2.5cm]{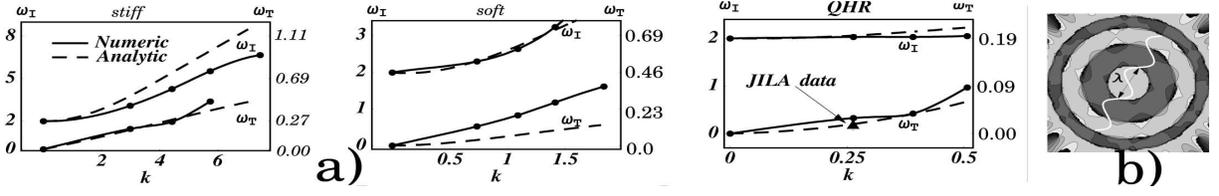}
\caption{a) The dispersion curves of azimuthally symmetric Tkachenko and inertial waves for 
lattices in different physical regimes. For each plot the scales labelled $\omega_{I}$ and 
$\omega_{T}$ on left and right ordinate are for inertial and Tkachenko modes, respectively. 
The wave vector for each plot is in units of $2\Omega/s$, where s = speed of sound at the trap centre.
Our QHR lattice is specified by $\mu / 2 \Omega=0.7 $. Triangle size reflects error bars. 
b) Extraction of the effective wavelength ($\lambda$) for both branches
of an azimuthally symmetric mode from the phase of the density fluctuation ($\rho$) of the bulk mode. 
In all phase plots we interpolate the bulk variations over the vortex cores and the 
phase changes from $0$ to $2 \pi$ as we move from dark to light.}
\label{profile}
\end{figure*}

Our methods~\cite{lobaks} are based on the Bogoliubov-deGennes mean field theory~\cite{svid} in a pancake trap geometry. The ground state in a rotating frame is obtained from the Gross-Pitaevski functional
$H= \int d^{2}r \left[\psi^{*}
\left(\hat{h}_{\bot} -\Omega \hat{L}_{z}-\mu\right)\psi+\frac{1}{2} g_{2D} N
|\psi|^{4}\right],$
where $\hat{h}_{\bot}= -\frac{\hbar^2}{2m} \nabla_{\bot}^2+\frac{1}{2}m \omega_{\bot}^2 {\bf r^2}$
is the single particle Hamiltonian in the $x-y$ plane.
We take $\hat{L}_{z}$ as the angular 
momentum operator defined in polar $(r,\phi)$ cordinates as $-i \hbar \frac{\partial }{\partial \phi}$. 
The quasi-two dimensional coupling constant $g_{2D} \equiv 2\pi g_{3D}/a_{z}$ where 
$g_{3D}=(4\pi a \hbar^2)/m$ and $a_{z}$ is the oscillator length along the $z$-axis:
$a_{z}=\sqrt{\hbar /m\omega_{z}}$. Unless stated otherwise we express energy, 
length and time in oscillator units: $\hbar\omega_{\bot}$, $\sqrt{\hbar /m\omega_{\bot}}$
and $1/\omega_{\bot}$ respectively. Writing the mean field fluctuation as 
$\delta\Phi_{o}=u({\bf r})e^{-i\omega t}-v({\bf r})^{*} e^{i\omega t}$, 
a first order expansion around the groundstate $\Phi_{o}$ yields the coupled equations for the 
Bogoliubov amplitudes $u(\bf r)$ and $v(\bf r)$: 
\begin{eqnarray}
(\hat{h}_{\bot}-\Omega \hat{L}_{z}+2g_{2D}N|\Phi_{o}|^{2}-\mu)v_{j}-gN(\Phi_{o}^{*})^{2}u_{j}
=-\omega_{j}v_{j} \nonumber \\
(\hat{h}_{\bot}-\Omega \hat{L}_{z}+2g_{2D}N|\Phi_{o}|^{2}-\mu)u_{j}-gN\Phi_{o}^{2}v_{j}=\omega_{j}u_{j} 
\label{coupled}
\end{eqnarray}
\begin{figure}
\includegraphics*[width=7.0cm,height=2.0cm]{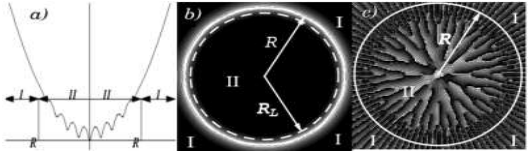}
\caption{ a) Cross section (along z-axis) of the effective mean field potential 
($\frac{1}{2}m \omega_{\bot}^2 {\bf r^2}+2gN|\Phi_{o}|^{2}-\mu$) experienced by a 
particle in the trap b) The phase and c) amplitude of the Bogoliubov quasi-particle state ($u$) 
for this excitation. $R_{L}= \langle r \rangle$ (dashed circle) 
for a state in the Landau state 
$h_{0,m}(r,\phi)$ where m is chosen to match that of u in region I.}
\label{surface_wave}
\end{figure}  
We solve these equations numerically.  In Fig.~\ref{profile}, we plot resulting dispersion curves, for parameters relevant to the recently measured~\cite{jilat,jilat2} azimuthally symmetric $(p,0)$~\cite{anglin,lobaks} modes in the three physical regimes described above. Here $p$ is the radial order of the excitation. 
As shown in (Fig.~\ref{profile}b), we can obtain the effective wave length ($\lambda$) for the excitation from a phase plot of the complex density fluctuation amplitude for the $j$th excitation $\rho_{j}\equiv \Phi_{o}^{*}u_{j}-\Phi_{o} v_{j}$.  Apart from quantum fluctuations, 
the real part of $\rho$ has the same meaning as the 
hydrodynamic density fluctuation~\cite{svid}. Consider the two scales labeled 
$\omega_{T}$ and $\omega_{I}$ for the branches in each plot. 
In accordance with hydrodynamic predictions~\cite{baym}, the reduction in $C_{1}$ and $C_{2}$ as the QHR is 
approached causes the inertial and Tkachenko waves to occur within increasingly smaller ranges in $\omega$ around
$\omega=0$ and $\omega=2$.  The triangular data point in Fig.~\ref{profile} represents recent JILA data~\cite{jilat,jilat2}, in excellent agreement
with this and related work~\cite{baym,lobaks}.

\begin{figure*}
\includegraphics*[width=15cm, height=7.5cm]{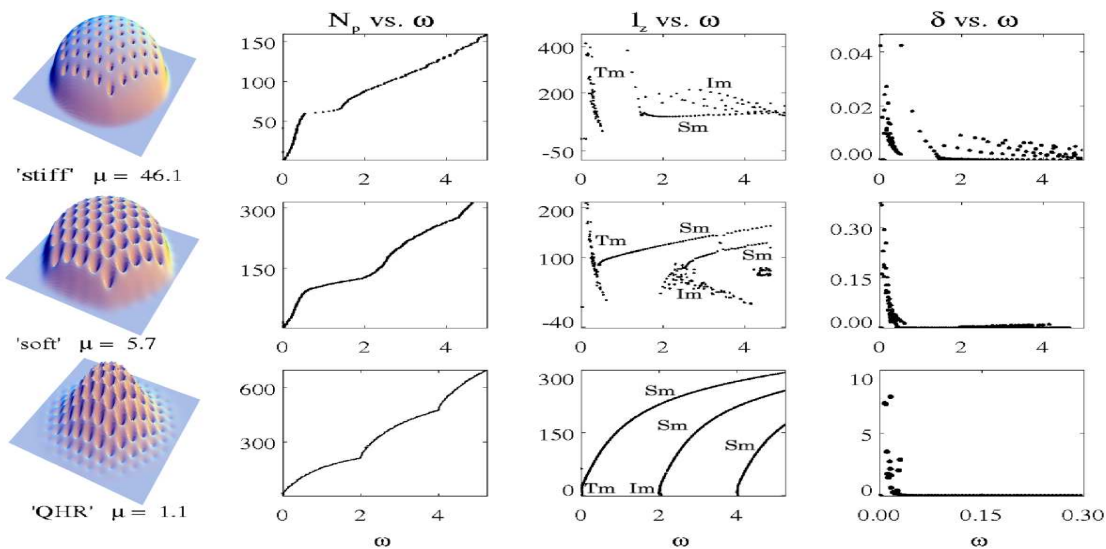}
\caption{We plot the $N_{p}$, $l_{z}$ and $\delta$ defined in the text for different 
ground states. The labels Tm, Im, Sm approximately mark the locations of
Tkachenko, inertial and surface modes respectively. The depletion is multiplied by a factor of $10000/N$.  Note the Gaussian shape of the cloud in the QHR.}
\label{khan}
\end{figure*}

To introduce the surface wave excitations -- the BEC edge states -- we employ the analogy between 
the BEC and an electron confined in 
a magnetic field~\cite{ho}. We will see that the quasi-particle states ($u$) responsible for 
these surface waves map onto single particle states of the Landau level picture.
First we rewrite the Hamiltonian as 
$\hat{h}_{\bot}-\Omega L_{z} = \hat{H}_{L}+\hbar(\omega_{\bot}-\Omega)\hat{L}_{z}$ where
$\hat{H}_{L}=\left(-i\hbar\nabla_{\bot}-m\omega_{\bot}\hat{\bf{z}} \times \bf{r}\right)^{2}/(2 m)$.
$\hat{H}_{L}$ is equivalent to the Hamiltonian of an electron of charge $e$ and 
mass $m$ moving in the $x-y$ plane under the influence of a magnetic
field $B \hat{\bf{z}}$ with vector potential $ A = \frac{1}{2} B \hat{\bf{z}} \times \bf{r}$. We then make the identification $eB/2m= \omega_{\bot}$. If we denote
the eigenfunctions of $\hat{H}_{L}$ as $h_{n,m}$ 
with eigenvalues $\epsilon_{n,m}=\hbar \omega_{\bot}(2n+1)$, then 
the integers $n$ and $m$ represent the Landau level and the angular momentum quantum numbers respectively.  The eigenvalue of $\hat{h}_{\bot}-\Omega \hat{L}_{z}$ may be written as $\omega_{n,m}=\hbar \left( (\omega_{\bot}+\Omega)n+(\omega_{\bot}-\Omega)m+\omega_{\bot}\right)$. Note that the positive term $\left(\omega_{\bot}-\Omega \right)m$ unilaterally lifts the degeneracy of the Landau levels and represents only
a small perturbation in the rapidly rotating limit $\Omega \approx \omega_{\bot}$.
To explore the analogy, we make two observations: First, the high angular momentum of these states ensures that 
$|\vec{k}|\xi \gg 1$ such that the excitation is particle-like; that is
that $ \left| v \right| \ll \left| u \right| $ implying that the properties of the excitation are dictated by the character of $u$. We may separately solve for $u$ in the regions $r \ge R$ and $r < R$ (labeled I and II in Fig.~\ref{surface_wave}) from Eq.~\ref{coupled} (top), ignoring the coupling.
Second, the large momentum of these states confines $|u|$ mostly to region I, near the condensate surface. There, the effective potential is harmonic because the interaction term $2g_{2D}N|\Phi_{o}|^{2}$ is negligible. Thus we find that the excitation coincides with the solutions of $\hat{H}_{L}$ represented by $h_{n,m}$. To visualize the edge states, we show in Figs.~\ref{surface_wave}b,c a full (no approximation) numerical solution of $\left| u \right|$ for an example. We see that the amplitude is radially peaked near $R_{L}=\sqrt{2 m} l$, with a spread of order $l=1/\sqrt{\Omega}$. This is exactly as would be expected for the equivalent Landau state defined by $h_{0,m}= \frac{1}{\sqrt{2 \pi 2^{m} m !} l} r^{m}e^{i m \phi}e^{-r^{2}/4 l^{2}}$
~\cite{yoshi}. We stress that these edge states cannot be adequately accounted for in any long wavelength treatment. Further, they cannot be neglected as they are bound to occur throughout the spectrum, and in all three regimes. 

Having identified the edge states, we turn to the spectrum evolution across the three regimes. Our results are summarized in Fig.~\ref{khan}. We calculate the total 
excitation number $N_{p}$ which is related to the density of states $\chi$ by 
$N_{p}(\omega)=\int_{0}^{\omega}\! \chi(\omega')d\omega'$. 
The second order contribution to the angular momentum 
$l_{z}=\int (u_{i}^*\hat{L}_{z} \,u_{i} -
v_{i}^*\hat{L}_{z}\,v_{i}) d{\bf r}$ is due to the density and velocity field fluctuations, and $\delta_{i} = \int\! \left| v_{i}({\bf r}) \right|^2 d{\bf r}$ is the condensate depletion of the $i$th excitation at $T\!=\!0$. By correlating these plots with MPEG movies of each excitation, we can assign a mode character to each individual spectral point. 

Consider the spectrum in the stiff regime. Particle-like excitations occur in the $l_{z}$ plots as families of points that organize into distinct lines with positive slope (labeled by $Sm$). By contrast, the more scattered points correspond to Tkachenko waves ($Tm$) in the interval $0 \leq \omega < 1$ and to inertial modes ($Im$) in the interval $\omega \geq 2$. As we move to the soft and then QHR, the $l_{z}$ spectrum is increasingly dominated by particle-like excitations mainly for two reasons. First, the associated drastic decrease in $C_{1}$ and $C_{2}$ with increasing $\Omega$ affects the mode energies and pushes the Tkachenko and inertial modes into smaller intervals of $\omega$ around $\omega=0$ and $\omega=2$ respectively (see also Fig.~\ref{profile}a). For the same reason, these modes experience a reduction in their speeds and are 
shifted towards smaller $l_{z}$. Second, what is key is that as $\Omega \rightarrow \omega_{\bot}$, $|\vec{k}|\xi \rightarrow \infty$ and inertial modes become increasingly organized and acquire particle-like character such that, deep in the QHR, they are well described by a superposition of closely spaced (single-particle) Landau level states $h_{n,m}$. 
\begin{figure}
\includegraphics*[width=8cm, height=1.7cm]{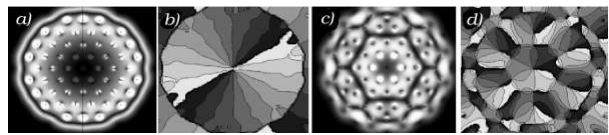}
\caption{ a) Amplitude and b) phase of $\rho$ for an inertial wave of long wavelength. 
c) Amplitude and d) phase of $\rho$ for an inertial wave of shorter wavelength. }
\label{lineup}
\end{figure}

The progression from collective to single particle character is well illustrated by the behavior of the amplitude and phase of $\rho$ for $\Omega \rightarrow \omega_{\bot}$ as shown in Figs.~\ref{lineup}. The mode in Figs.~\ref{lineup}a,b is consistent with a hydrodynamic approximation~\cite{baym,cozzini,choi} which stipulates a functional 
form $\rho=f(r)e^{ip \phi}$, where $p$ is an integer. As $\Omega$ increases and the lattice becomes softer, the approximation breaks down for increasingly longer wavelengths and features signalling particle character begin to appear as interference structures on the scale of $\epsilon$ (the `islands' in Figs.~\ref{lineup}c,d). 
\begin{figure}
\includegraphics*[width=8cm, height=1.7cm]{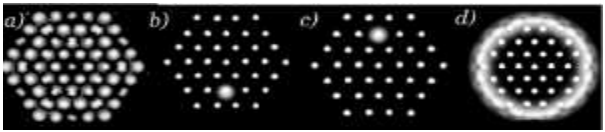}
\caption{ a) Amplitude of $\rho$ for a Tkachenko wave. b) and  c) represent 
$|\psi|$ for degenerate bound Hartree-Fock states, where $\psi$ is the wavefunction. d) 
$|\psi|$ of a extended Hartree-Fock state. In all plots the overlayed white dots are the equilibrium 
vortex positions}
\label{particle}
\end{figure}
For particle-like excitations, $l_z$ has a similar meaning to $m$ 
for the $h_{n,m}$ states. We may thereby
 interpret the structure of the particle-like sub-spectrum from the 
expression for $\omega_{n,m}$. If we identify the groundstate energy $\mu$ with the zero point energy ($\hbar\omega_{\bot}$) of $\omega_{n,m}$ then the eigenvalue for $\hat{h}_{\bot}-\hat{L}_{z}\Omega$ 
suggests bands centered at 
$\omega=2n: n\in\left\{0,1,2,..\right\}$. Physically, the large $m$ of the surface states
ensures that they occupy larger intervals of the $\omega$ axis around the Landau levels 
than inertial or Tkachenko modes due to the term $\left(\omega_{\bot}-\Omega \right)m$. Thus, 
the whole spectrum acquires a band-like structure as seen in the $N_{p}$ plots of Fig.~\ref{khan}. 

Having identified the connection of the BEC surface waves to IQHE edge states, 
a casual glance at $|\rho|$ for a Tkachenko (Fig.~\ref{particle}a) and 
an inertial mode (Fig.~\ref{lineup}a) suggests an extension of the BEC analogy to other single particle states of the FET: 
The Tkachenko waves, confined to the vortex cores, are analogues of 
the bound states and the inertial modes, which cover the entire array (see Fig.~\ref{lineup}b), are analogues of the extended states. Conceptually, the vortex cores play the role of charged impurities in the FET. However one 
important difference is that in a BEC, the Tkachenko 
modes are generally collective and not particle-like due to
significant mean field dynamics. To explore the connection we solve the related Hartree-Fock problem defined by Eq.~\ref{coupled} minus the coupling term. This is equivalent to assuming a nearly incompressible lattice, i.e. to infinite values of $C_{1}$ and $C_{2}$. In addition to surface excitations, we obtain both a group of degenerate single 
particle states bound to the vortex cores (Fig.~\ref{particle}b,c) as well as extended states (Fig.~\ref{particle}d) which cover the array. As the $C_{1}$ and $C_{2}$ acquire finite values, these states map back onto the Tkachenko and inertial modes of the full problem and have lower energies than their Hartree-Fock counterparts.  

Consider next the condensate depletion $\delta$. As we approach the QHR, the density drops, $|\vec{k}|\xi \rightarrow \infty$ and $|\vec{k}|\epsilon \rightarrow 0$. The inertial modes become particle-like and contribute less to $\delta$, while the Tkachenko waves become more 'collective' and contribute more to $\delta$. Moreover, the total 
depletion ($\sum_{i}\! \delta_{i}$) grows. This can be seen in Fig.~\ref{khan}.  In concurrence with work predicting a quantum phase 
transition~\cite{sinova,baym_modes} as $\Omega \rightarrow \omega_{\bot}$, we see
in Fig.~\ref{khan},  a significant increase in the depletion of the Tkachenko modes as $\Omega \rightarrow \omega_{\bot}$. 
Like phonon vibrations in a crystal lattice, the 
Tkachenko waves arise out of a broken continuous symmetry, and 
are a path to the destruction of the array and, in this case, destruction of the condensate as thermal and/or quantum fluctuations grow.

The IQHE framework for describing the excitations can be used to understand many properties of the rotating BEC. For example, vortex nucleation. A small increase in $\Omega$ is equivalent in the electron picture to an increase in the magnetic field strength. As B increases, the radius of the classical orbital `ring' is reduced to keep the flux enclosed by the degenerate
electron orbit constant. In the BEC, the decrease in the edge state radius with increasing $\Omega$ causes vortices of the precessing ring to be drawn towards and included into the interior of the condensate. This evolution was detected in recent models in which vortices were observed to enter into the bulk of the condensate via a precessing
ring~\cite{gardiner}. Quantitatively, we consider a perturbation of Eqs.\ref{coupled} by $\Omega \rightarrow \Omega + \delta\Omega$ starting from $\Omega \sim 0$. 
We find that the lowest edge state acquires a negative energy, falling below the ground state -- a signature of a thermodynamic instability.
This mechanism is particularly relevant to the experimental method pioneered at 
JILA~\cite{jilat2} in which the condensate is nucleated from a rotating thermal cloud. 

In sum, the Landau level analogy provides an organizing principle for vortex lattice investigations in rotating BECs. In addition to describing the particle-like excitations of the lattice which grow in number as $\xi \rightarrow \infty$, it inspires further analogies with single particle states of the FET.

We thank Michael Banks for computing assistance.
This work is supported by NSF, ARO and ONR. Insightful discussions with H. Pu, 
C. K. Law and G. S. Krishnaswami are acknowledged.     


\begin{thebibliography}{99}
\bibitem{sonin} E. B Sonin, Rev. Mod. Phys. {\bf 59}, 87 (1987). 
\bibitem{anglin} J. Anglin and M. Crescimanno, e-print cond-mat/0210063.
\bibitem{baym} G. Baym, e-print cond-mat/0305294. 
\bibitem{lobaks}L. O. Baksmaty, S. J. Woo, S. Choi and N. P. Bigelow, e-print cond-mat/0307368.
\bibitem{nippon}T. Mizushima, Y. Kawaguchi, K. Machida, 
T. Ohmi, T. Isoshima and M. M. Salomaa, e-print cond-mat/0308010.
\bibitem{cozzini}M. Cozzini and S. Stringari, Phys. Rev. A 67, 041602 (1998). 
\bibitem{choi}S. Choi, L. O. Baksmaty, S. J. Woo and N. P. Bigelow, e-print cond-mat/0306549. 
\bibitem{jilat} I. Coddington, P. Engels, V. Schweikhard and E.A. Cornell, e-print cond-mat/0305008.
\bibitem{jilat2} V. Schweikhard, I. Coddington, P. Engels 
, V.P. Morgendorff and E.A. Cornell e-print cond-mat/0308582.
\bibitem{yoshi} See, D. Yoshioka, {\em The quantum Hall effect} (Springer-Verlag, Berlin, 2002).
.
\bibitem{ho}Tin-Lun Ho, Phys. Rev. Lett. {\bf 87}, 060403 (2001).
\bibitem{wilkin}N. R. Cooper, N. K. Wilkin, J. M. F. Gunn, Phys. Rev. Lett. {\bf 87}, 120405 (2001).
\bibitem{svid}A. L. Fetter and A. A. Svidzinsky, J. Phys. Condens. Matter {\bf 13}, R135 (2001). 
\bibitem{gardiner}A. A. Penckwitt, R. J. Ballagh, C. W. Gardiner, Phys. Rev. Lett. {\bf 89}, 260402 (2002).
\bibitem{sinova}J. Sinova, C.B. Hanna,A. H. MacDonald, Phys. Rev. Lett. {\bf 89}, 030403 (2002).
\bibitem{baym_modes} G. Baym, e-print cond-mat/0308342.
\end{thebibliography}
\end{document}